\documentclass{article}

\usepackage{geometry}
\usepackage{amssymb, amsmath, systeme}
\usepackage{graphicx}
\usepackage[hidelinks]{hyperref}
\usepackage{natbib}
\usepackage{bm}
\usepackage[font=small,labelfont=bf]{caption}

\mathchardef \hyp = "2D

\graphicspath{{./}}

\title{An approach for identifying sources of inadequacy and upgrades in models with high-dimensional outputs and boundary conditions}
\author{Filippo Monari\\
 \small Mechanical and Aerospace Engineering, University of Strathclyde, Glasgow (UK)\\
 \small \texttt{filippo.monari@strath.ac.uk}}

\begin{document}

\maketitle

\begin{abstract}
  The construction of computer models (mathematical models implemented in computer codes), with respect to observed phenomena, is usually undertaken
  by building different variants depending on modeller sensibility, and choosing the one yielding the best fit of the field data, according to Root
  Mean Squared Error (RMSE) based measures. Usually a particular model is chosen because of its marginally lower RMSE, and not because of its actual
  higher adequacy, risking that its capability of extrapolating predictions is poor. This work aims at improving the current practice in the creation
  of computer models by proposing an approach similar to those employed in statistical modelling, wherein starting from the simplest hypothesis,
  effective model upgrades are identified by analysing discrepancies between observations and predictions, and different model variants are compared
  according to robust likelihood based criteria. The method, focused on models with high dimensional outputs and boundary conditions and centred on
  Bayesian calibration, is demonstrated on numerical experiments considering a series of building energy models. The object of the modelling is a test
  facility used for round robin tests in the context of the International Energy Agency (IEA), Energy Building and Communities (EBC), Annex 58.
\end{abstract}

\section{Introduction}
Mathematical models implemented in computer codes (computer models) have reached great capabilities at representing real world phenomena, and nowadays
are indispensable for solving many problems related to different fields of engineering. Nonetheless, they are characterised by complex mathematical
structures, and their creation requires a significant amount of detailed information (prior information), which is not always available and usually
affected by substantial uncertainties. Often, it is necessary to tailor the model structure according to the available prior information and to its
embedded uncertainties, by making assumptions, deciding which aspects deserve more focus, and selecting suitable abstractions. It is useful to make an
example, connecting with the case study presented later, to better clarify this point. With respect to building energy models, which are perfect
examples of complex computer models with many inbuilt integrated components attempting to represent reality, even modelling the simplest building
involve choosing among several possibilities, for example: which materials to assume for the envelop components, whether or not to consider air
infiltration, in the latter case whether to adopt a constant infiltration model or a wind driven infiltration model, and whether to use an ideal
abstraction for the heating/cooling system rather than a more detailed dedicated model.

With respect to the modelling of observed processes, the development of computer models involves (i) the treatment and reduction of the uncertainties
contained within the prior information, and (ii) the assessment of the adequacy of the proposed model structure, that is the adequate consideration of
all the processes underpinning a the real phenomenon being modelled. While the former issue has been object of substantial research, resulting in the
formulation of methodologies such as Bayesian calibration of which a brief literature review follows, the latter has been rarely investigated
\citep{Strong2012, Strong2014}.

In practice, suitable model structures are defined by assessing the capabilities of matching field data of different variants, built depending on
modeller sensibility and expertise, according to metrics based upon the Root Mean Squared Error (RMSE). Because of its discretionary character, the
overparametrisation caused by the complex underpinning mathematical structures, and equifinality (i.e. different models yielding very similar RMSE,
despite having significantly different characteristics), this approach often fails in selecting the most suitable model of the observed
phenomenon. Indeed, a particular model structure is chosen just because of its marginally better RMSE, and not Because of its actual higher adequacy,
risking that its capability of extrapolating predictions is poor. This work aims at improving the current practice for the creation of computer models
by proposing an approach similar to those employed in statistical modelling, wherein starting from the simplest hypothesis, effective model upgrades
are identified by analysing discrepancies between observations and predictions, and different model variant are compared according to likelihood based
criteria (e.g. Akaike Information Criterion, Bayes Information Criterion or Likelihood Ratios) resilient to overparameterisation and equifinality. The
presented approach focuses on models with high dimensional outputs and boundary conditions, and is centred on Bayesian calibration.

Bayesian calibration is a popular framework used to infer parameter values in computer models. The first mathematical formulation was described in
\cite{Kennedy2001}. Building on this formulation, a series of works have demonstrated the capability of Bayesian calibration in a number of different
case studies \citep{Higdon2004, Bayarri2005, Bayarri2005a}. Subsequently, the representation of high-dimensional model outputs through basis
expansions has allowed the application of the Bayesian paradigm to the calibration of a wide range of computer models \citep{Heitmann2006,
  Bayarri2007, Higdon2008}. Generally Bayesian calibration relies on probabilistic models based on Gaussian Process Regression \citep{Rasmussen2006},
allowing the Bayesian inference of the computer model parameters (calibration parameters): the simulation model, providing a probabilistic emulator of
the computer model, and the observation model, giving a probabilistic depiction of the observed data. The latter in turn is composed of the
calibration model, allowing the estimation of the calibration parameters, and the discrepancy model, which represent the difference between model
predictions and observations due to inadequacies within the computer model (discrepancy). Identifiability problems may raise due to interference
between the discrepancy model, and the calibration model \citep{Kennedy2001, Bayarri2005}. These issues have been object of recent research, and
possible solutions have been proposed in \cite{Plumlee2017, tuo2015}.

In the presented approach, as in \cite{Strong2012, Strong2014}, the identification of deficiencies in the model structure is undertaken through the
modelling of the discrepancy between predictions and observations. These studies propose the assessment of the model adequacy, by modelling the
discrepancy within the mathematical structure of the model, which is decomposed in a series of sub-functions. Since the mathematical structure of
computer models is not usually directly accessible, in this work the discrepancy is modelled externally by exploiting the discrepancy term included in
the Bayesian calibration framework. In particular, a Gaussian process prior probability density distribution that avoids conflicts with the simulation
and calibration counterparts is adopted. Its hyper parameters are linked to the boundary conditions and used to quantify their correlation with the
discrepancy between predictions and measurements. This information can then be used by the modeller for proposing model upgrades, directly addressing
those aspects affected by the boundary conditions with higher correlations. Different models are compared by means of Bayes Factors. The proposed
methodology is demonstrated on numerical experiments considering synthetically generated and measured data, which analyse a series of building energy
models. The modelling environment is provided by the dynamic simulation program ESP-r \citep{Clarke2001a}. The object of the modelling is a test
facility used for round robin tests in the context of the International Energy Agency (IEA), Energy Building and Communities (EBC), Annex 58
(\url{http://www.iea-ebc.org/projects/completed-projects/ebc-annex-58}). The aim of this international project was to investigate the capability of
different methods for characterising thermal properties and features of building envelopes, of which the test box was a simplified, but realistic,
surrogate.

\section{Method}
The addressed problem is the following.  A real process, subject to $S$ known time varying boundary conditions
($\bm X = [\bm x_1,\dots,\bm x_s,\dots,\bm x_S]$), and determined by unknown time independent parameters, is observed by measuring $N$ times a target
variable ($\bm y^*$). $R$ different models of the observed process, accounting for $\bm X$, are available.  The $r\hyp th$ model ($\mathcal M_r$) has
unknown underlying mathematical structure, and $P$ free parameters ($\bm z = [z_1,\dots,z_p,\dots,z_P]$) representing the time independent unknowns
($\bm z^*=[z^*_1, \dots, z^*_p, \dots, z^*_P$).  Having a sample of $M$ $\bm z$ ($\bm Z = [\bm z_1,\dots,\bm z_m,\dots,\bm z_M]^T$), $\mathcal M_k$
provides a set of predictions for $\bm y^*$ ($\bm Y = [\bm y_1,\dots,\bm y_m,\dots,\bm y_M]$). Therefore, it is necessary to: (i) infer the values of
the parameters in $\bm z^*$ (calibration parameters); (ii) learn about causes of discrepancy between predictions and measurements, in order to provide
effective model improvements; (iii) effectively select among the $R$ created variants the model providing the best description of the observed
process. To solve these problems, the following three steps are undertaken: (i) quasi-Bayesian calibration, (ii) discrepancy analysis, and (iii) model
selection. Their explanation follows.

\subsection{Quasi-Bayesian calibration}\label{S:qBayes_cal}
This section describes the mathematical framework used for model calibration, which builds on \cite{Higdon2008}.  However, the involved probabilistic
models are explicitly expressed as a factorisation of terms, which are easier to process. The overall framework is quasi-Bayesian, that is the hyper
parameters of the simulation model are optimised and fixed afterwards. This approximation is generally accepted since, while giving results similar to
a fully Bayesian analysis, it allows a great simplification of the calculation involved \citep{Bayarri2007}.

\subsubsection{Simulation model}\label{S:simMdl}
Each simulation output ($\bm y_m$) is represented by the sum of the model emulator ($f(\bm z_m) = \bm Kw(\bm z_m)$) and white noise
($\bm \varepsilon \sim \mathcal N(0, \lambda^{-1})$, with $\lambda \sim \mathcal Gamma(a, b)$).
\begin{equation}
  \bm y_m = f(\bm z_m) + \bm \varepsilon = \bm Kw(\bm z_m) + \bm \varepsilon
  \label{EQ:sim_model__1}
\end{equation}
where the $N \times Q$ matrix $\bm K$ has as columns orthogonal vectors ($\bm k_q$) defined as explained in Section \ref{S:bases}.  The function
$w(\cdot)$ can be expressed with $Q$ Gaussian Processes ($\mathcal{GP}$s), modelling the $Q$ set of coefficients associated with the vectors
$\bm k_q$:
\begin{equation}
  w_q(\bm Z_q) \sim \mathcal{GP}(\bm 0, \varrho_q(\bm Z_q, \bm Z_q, \bm \gamma_q) + \bm I\lambda_q^{-1})
  \label{EQ:sim_model__2}
\end{equation}
where $\bm \gamma_q$ represents the hyper parameters, and $\bm Z_q = [\bm z_{q,1},\dots,\bm z_{q,m},\dots,\bm z_{q,M}]^T$, with $\bm z_{q,m}$ the
vector containing the subset of values of $\bm z_m$ relative only to the parameters significant for the $q\hyp th$ $\mathcal{GP}$. In particular these
parameters are selected according to the forward selection strategy outlined in \cite{Welch1992}.

Optimal values of the precision parameters ($\bm \lambda = [\lambda_1,\dots,\lambda_q,\dots,\lambda_Q]$), and of the covariance function hyper
parameters ($\bm{\gamma} = [\bm{\gamma}_1,\dots,\bm{\gamma}_q,\dots,\bm{\gamma}_Q]$), are determined by maximising the corresponding joint probability
density distribution conditional on $\bm W = [\hat{\bm w}_1, \dots, \hat{\bm w}_q, \dots, \hat{\bm w}_Q]$ and $\bm Z$:
\begin{eqnarray}
  && p(\bm \lambda, \bm \gamma | \bm W, \bm Z) \propto \nonumber\\ 
  && \prod_{q=1}^{Q} [(2\pi)^{-\frac{M}{2}}|\bm I\lambda_q^{-1} +  \varrho_q(\bm Z_q, \bm Z_q, \bm \gamma_q)|^{-\frac{1}{2}} exp\{-\frac{1}{2} \hat{\bm w}^T_q
     [\bm I\lambda_q^{-1} +  \varrho_q(\bm Z_q, \bm Z_q, \bm \gamma_q)]^{-1} \hat{\bm w}_q \} \times \nonumber\\
  && \times \frac{(\frac{b^\prime}{\bm k_q^T\bm k_q})^{a^\prime}}{\Gamma(a^\prime)}\lambda_q^{a^\prime - 1}
     exp\{-\frac{b^\prime}{\bm k_q^T\bm k_q} \lambda_q\} \times p(\bm \gamma_q)]
     \label{EQ:sim_model__4}
\end{eqnarray}
where:
\begin{eqnarray*}
  a^\prime &=& a + \frac{M(N-Q)}{2} \label{EQ:sim_model__3.1}\\
  b^\prime &=& b + \frac{1}{2}\sum_{m=1}^{M} \bm y_m^T (\bm I - \bm K (\bm K^T \bm K)^{-1} \bm K^T) \bm y_m\\
  \hat{\bm w}_q &=& (\bm k_q^T\bm k_q)^{-1}\bm k_q^T \bm Y
\end{eqnarray*}

The $r\hyp th$ model, probabilistically characterised by the calculated optimal values of the corresponding $\bm{\gamma}$ and $\bm \lambda$
parameters, is indicated with the symbol $\mathcal M_r$.

\subsubsection{Observation model}
The model representing the measured observation ($\bm y^*$) is composed of: (i) the calibration model, that is the physical model emulator evaluated
at the unknown parameter values $\bm z^*$ ($f(\bm z^*) = \bm K w^*(\bm z^*)$); (ii) the discrepancy model, dependent only on the known variable
boundary conditions ($\Delta(\bm X) = \bm H v(\bm X)$); and (iii) i.i.d. Gaussian noise ($\bm \varepsilon^* \sim \mathcal N(0, \lambda^{*^{-1}})$,
with $\lambda^* \sim \mathcal Gamma(a^*, b^*)$):
\begin{equation}
  \bm y^* = f(\bm z^*) + \Delta(\bm X) + \bm \varepsilon^* = \bm Kw^*(\bm z^*) + \bm H v(\bm X) + \bm \varepsilon^*
  \label{EQ:obs_mdl__0}
\end{equation}
where $\bm H$ is an $N \times (N-Q)$ matrix, having as columns orthonormal vectors ($\bm h_d$) orthogonal to $\bm K$ (Section \ref{S:bases}), and
$v(\bm X) \sim \mathcal{GP}(\bm 0, \bm H^T\zeta(\bm X)\bm H)$.  Thus, $\bm y^*$ is decomposed according to two models, $f(\bm z^*)$ and
$\Delta(\bm X)$, spanning complementary spaces, defined by $\bm K$ and $\bm H$ respectively.  The former explains the variance of $\bm y^*$ lying
within the simulation space, while the latter explains the variance of $\bm y^*$ lying outside (or orthogonal to) such space (i.e. the discrepancy).

The calibration model, is used to estimate the calibration parameters ($\bm z^*$), and it is developed as follows.  Analogously to Section
\ref{S:simMdl}, the function $w^*(\cdot)$ is expressed by $Q$ $\mathcal{GP}$s.  However, in this case the $\mathcal{GP}$s are conditioned upon their
counterparts in Equation (\ref{EQ:sim_model__2}):
\begin{equation*}
  w_q^*(\bm z_q^*) \sim \mathcal{GP}(w^\prime_q, \sigma^{\prime^2}_q) 
\end{equation*}
where:
\begin{eqnarray*}
  w_q^\prime &=& \varrho_q(\bm z_q^*, \bm Z_q, \bm \gamma_q)[\varrho_q(\bm Z_q, \bm Z_q, \bm \gamma_q) + \bm I\lambda_q^{-1}]^{-1}\hat w_q \\
  \sigma^{\prime2}_q &=& [\lambda^{*-1}_q + \varrho_q(\bm z_q^*,\bm z_q^*, \bm \gamma_q)] - \varrho(\bm z_q^*, \bm Z_q)[\varrho_q(\bm Z_q, \bm 
                         Z_q, \bm \gamma_q) + \bm I\lambda_q^{-1}]^{-1}\varrho(\bm Z_q,\bm z_q^*, \bm \gamma_q) 
\end{eqnarray*}

Posterior probability density distributions for the calibration ($\bm z^* = \bigcup \{\forall \bm z_q : q = 1, ..., Q\}$) and
precision($\bm \lambda^*=[\lambda_1,\dots,\lambda^*_q,\dots,\lambda^*_Q]$) parameters are inferred by integrating with Markov Chain Monte Carlo (MCMC)
methods their joint posterior probability density distribution, given $\bm{\hat w}^*=[\hat w^*_1,\dots,\hat w^*_q,\dots,\hat w^*_Q]$ (where
$\hat{w}^*_q = (\bm k^T_q \bm k_q)^{-1} \bm k^T_q \bm y^*$), and $\mathcal M_r$:
\begin{eqnarray}
  && p(\bm z^*, \bm \lambda^* | \bm{\hat w}^*, \mathcal M_r) \propto \nonumber\\
  && \prod_{q=1}^Q [(2\pi\sigma_q^{\prime2})^{-\frac{1}{2}} exp\{-\frac{(\hat{w}_q^* - w_q^\prime)^2}{2\sigma_q^{\prime^2}}\} \times 
     \frac{(\frac{b^*}{\bm k_q^T\bm k_q})^{a^*}(\lambda^*_q)^{a^* - 1}}{\Gamma(a^*)}
     exp\{-\frac{b^*\lambda^*_q}{\bm k_q^T\bm k_q}\}] \times p(\bm z^*) 
     \label{EQ:obs_mdl__1}
\end{eqnarray}
$p(\bm z^*)$ indicates the prior probability density distributions chosen for $\bm z^*$.

The discrepancy model ($\Delta(\bm X)$) is used to investigate correlation relationships between boundary conditions and the discrepancy (Section
\ref{S:diff_analysis}).  Empirical probability density distributions for its unknown hyper parameters and precision parameter $\lambda^*$, are
estimated by integrating with MCMC algorithms the relative joint posterior probability density distribution given
$\bm{\hat v} = (\bm H^T\bm H)^{-1}\bm H^T \bm y^*$, and $\bm X$:
\begin{equation}
  \begin{split}
    & p(\bm \gamma^*, \lambda^*|\bm{\hat v}, \bm X) \propto \\
    & (2\pi)^{-\frac{1}{2}}|\bm H^T(\lambda^{*-1} + \zeta(\bm X, X))\bm H|^{-\frac{1}{2}}
    exp\{-\frac{1}{2}\bm{\hat v}^T[\bm H^T(\lambda^{*-1} + \zeta(\bm X, \bm X))\bm H]^{-1}\bm{\hat v}\} \times \\
    & \frac{(a^{*\prime})^{b^{*\prime}}}{\Gamma(a^\prime)}(\lambda^*)^{a^{*\prime} - 1} exp\{-b^{*\prime} \lambda^*\} \times p(\bm z^*) \times
    p(\bm{\gamma}^*)
  \end{split}
  \label{EQ:obs_mdl__2}
\end{equation}

The MCMC method used to integrate Equations (\ref{EQ:obs_mdl__1}) and (\ref{EQ:obs_mdl__2}) is the Annealed Importance Sampling (AIS) algorithm
\citep{Neal2001}, which allows also the calculation of the respective normalising constants used for performing model selection (Section
\ref{S:model_selection}).

\subsubsection{Basis systems definition}\label{S:bases}
The field observations ($\bm y^*$) are decomposed according to the basis systems $\bm K$ and $\bm H$, the former spanning the model simulation space
and the latter spanning the complementary space needed to completely explain their variance.

The matrix $\bm K$ is defined as follows.  The first $Q-1$ columns are a subset of the eigenvectors of the matrix $\bm Y_c\bm Y_c^T$ (where $\bm Y_c$
is the matrix $\bm Y$ after having centred its columns on zero) scaled by the respective singular values.  In particular, these eigenvectors are
calculated by Singular Value Decomposition (SVD), and selected according to the relative eigenvalues, so as to explain the 99\% of the total variance
contained in $\bm Y_c$.  The $Q\hyp th$ column of $\bm K$ is the vector composed of all ones, so that $\hat{\bm w}_Q$ corresponds to the subtracted
simulation mean values.  In this way the vectors $\bm k_q$, besides being orthogonal, are also completely uncorrelated, which improves the
identifiability of the model parameters.

For the definition of $\bm H$ the approach explained below has been adopted.  Let $\bm P = \bm I - \bm K (\bm K^T\bm K)^{-1}\bm K^T$ be the matrix
defining the projection in the space orthogonal to that spanned by $\bm K$.  This matrix has $N-Q$ eigenvalues equal to one and $Q$ eigenvalues equal
to zero \citep{Hodges2010}.  The columns of the matrix $\bm H$ are then defined as the former. Therefore, since $\bm P = \bm H\bm H^T$, the
discrepancy model is built as projection of $\mathcal{GP}(\bm 0, \zeta(\bm X))$ in the space complementary to the one spanned by the calibration
model, so as to avoid confounding between them.

\subsection{Covariance functions and hyper parameters}\label{S:cf_hp}
The Equations (\ref{EQ:cf1}) and (\ref{EQ:cf2}) (wherein $\delta_{i,j}$ is the Kronecker delta) display the functions $\varrho(\cdot,\cdot)$ and
$\zeta(\cdot, \cdot)$, which are represented with square exponential kernels, but parametrised in a similar way as in \cite{Higdon2008}. A white
kernel is also added with respect to the former.
\begin{eqnarray}
  \varrho_q(\bm z_{i,q}, \bm z_{j,q}) &=& \frac{1 - \sigma^2_q}{\sigma_q^2} \prod_{p=1}^P \beta_{p,q}^{4(z_{i,p,q} - z_{j,p,q})^2} +
                                          \delta_{i,j}\frac{1 - \eta^2_q}{\eta^2_q} \label{EQ:cf1}\\
  \zeta(\bm x_i, \bm x_j) &=& \frac{1 - \tau^2}{\tau^2} \prod_{s=1}^{S} \alpha_s^{4(x_{s,i} - x_{s,j})^2} \label{EQ:cf2} 
\end{eqnarray}
Such parametrisation has been found convenient because it limits the hyper parameters space to a unit hypercube. In the following the prior
probability density distributions adopted for the hyper parameters in Equations (\ref{EQ:cf1}) and (\ref{EQ:cf2}) are explained, as well as the
underpinning rationales.

For the parameters $\sigma_q^2$, $\tau^2$, $\beta_{p,q}$ and $\eta^2_q$ uniform probability density distributions have been adopted. With respect to
the function $\varrho(\cdot, \cdot)$, the hyper parameters $\sigma_q^2$ govern the fraction of model output variance captured by the emulator, while
the hyper parameters $\eta^2_q$ represent explicitly the residual model output variance, due mainly to neglected parameters. Thanks to the property of
Automatic Relevance Determination (ARD) \citep{Neal1996}, naturally possessed by the $\mathcal{GP}$ models, they can easily be traded-off during the
optimisation of the emulator, effectively avoiding overfitting. ARD also act on the hyper parameters $\beta_{p,q}$, which will be pushed towards 1 if
the respective model parameters have little importance.

The hyper parameters $\alpha_s$ are linked to the boundary conditions ($\bm x_s$), of which they measure the correlation with the discrepancy
process. In this case, $\mathcal Beta(1, 0.1)$ probability density distributions have been chosen in order to strongly aid ARD, and heavily penalise
models having $\bm x_s$ correlated with their discrepancy processes. This set-up is particularly important in order to correctly perform model
selection. Indeed, by assuming uniformative prior probability density distributions for these hyper parameters, a highly inadequate model could have a
marginal likelihood similar to a more correct one, simply because its discrepancy model yields high likelihood values.

The shape ($a$) and rate ($b$) of the gamma probability density distribution used as prior for the simulation precision parameter ($\lambda$), were
set respectively to $2$ and a small value. For practical computational reasons, the latter was defined as the square root of the machine precision
times the highest singular value calculated through SVD. These choices had the objectives of yielding a probability density distribution not too
informative, while representing the virtually null amount of noise contained in the simulation outputs.

Similarly, the shape ($a^*$) and rate ($b^*$) of the gamma prior probability distribution used for the observation precision ($\lambda^*$) were set so
as to represent the prior information about measurement errors.  In particular, $a^* = N \cdot c$ and $b^* = var(\bm \nu) \cdot N \cdot c$, where
$\bm \nu$ are the observation errors, and $c \in [2/N, 1]$.  Thus $\lambda^*$ has mean equal to the precision of the measurement errors, and variance
inversely proportional to $c$, which can be interpreted as the confidence in the information used to determine the extent of these errors.

\subsection{Discrepancy analysis}\label{S:diff_analysis}
Discrepancy analysis consists of employing the discrepancy model to measure the correlation of each boundary condition ($\bm x_s$) with the
discrepancy. This correlation is quantified with the generalised correlation coefficient ($\mathcal R^2$) \citep{doksum1995}. The significance of a
particular $\bm x_s$ is established by comparing the relative $\mathcal R^2$ index with that corresponding to a fictitious variable $\bm x_0$
consisting of Gaussian i.i.d. noise, added in advance to the set of boundary conditions. This comparison is undertaken according to the procedure
outlined here below.

For a particular $\bm x_i$ the relative correlation coefficient $\mathcal R_i^2$ is evaluated as
\begin{equation*}
  \mathcal R^2_i = \frac{var[v(\bm X, \lambda^*, \tau, \alpha_i, \alpha_{j \neq i}=1)]}{var(\hat{\bm v})}
\end{equation*}
where
\begin{equation*}
  \begin{split}
    &v(\bm X, \lambda^*, \tau, \alpha_i, \alpha_{j\neq i}=1) = \\
    &\bm H^T \zeta(\bm X, \bm X, \tau, \alpha_i, \alpha_{j\neq i}=1) \bm H[\bm H^T (\bm I\lambda^* + \zeta(\bm X, \bm X, \tau, \alpha_i, \alpha_{j\neq
      i}=1) \bm H]^{-1}\hat{\bm v}
  \end{split}
\end{equation*}
$i$ and $j$ $\in [0, s : s = 1, \dots, S]$. In particular, due to the function chosen for $\zeta(\cdot)$, setting $\alpha_{j \neq i}=1$ makes
uninfluential all but the $i\hyp th$ boundary condition. Thus, $\mathcal R_i^2$ represents the fraction of discrepancy variance attributable to
$\bm x_i$ alone, given a particular value of the relative $\alpha_i$, $\lambda^*$ and $\tau$. For each $\mathcal R_i^2$ an empirical probability
density distribution is inferred from the posterior samples of the $\alpha_i$, $\lambda^*$ and $\tau$ parameters. Finally estimates and 95\% high
density intervals (HDI) for the quantity $\tilde{\mathcal R}^2_s = \mathcal R_s^2 - \mathcal R_0^2$ are calculated. The boundary conditions having 0
lying outside the HDI of their $\tilde{\mathcal R}_i^2$ are considered significant in determining the discrepancy between model prediction and field
measurements. Significant boundary conditions are ranked according to the respective $\tilde{\mathcal R}^2_i$ values.

The interpretation of these results can then be used to upgrade the model, by improving the representation of those processes influenced by
significant boundary conditions with high $\tilde{\mathcal R}^2$ indexes.

\subsection{Model selection}\label{S:model_selection}
Bayes Factors \citep{Kass1995} were adopted as the criteria for performing model comparison and selection. Given the observations $\bm y^*$, and two
competing models ($\mathcal M_i$ and $\mathcal M_j$), the Bayes Factor expressing the evidence supporting $\mathcal M_j$ against $\mathcal M_i$
($\mathcal B_{j,i}$) corresponds to the ratio between the probability that $\mathcal{M}_j$ generates $\bm y^*$ and the probability that
$\mathcal{M}_i$ generates $\bm y^*$:
\begin{equation*}
  \mathcal B_{j,i} = \frac{p(\bm y^*|\mathcal M_j)}{p(\bm y^*|\mathcal M_i)}
\end{equation*}
In this case, the probability that the $r\hyp th$ model ($\mathcal M_r$) generates $\bm y^*$ is proportional to the integral in Equation
(\ref{marginal_lkh}), wherein the subscript $r$ indicates the relation with model $\mathcal M_r$, and $\mathcal S$ indicates the space of the
parameters listed in its subscript.
\begin{equation}
  \begin{split}
    & L(\bm y^*| \mathcal M_r) = \\
    & \int_{\mathcal S_{\bm z^*_r, \bm \lambda_r^*}} p(\bm{\hat w}^* | \bm z^*_r, \bm \lambda^*_{r}, \mathcal M_r) p(\bm z^*_r) p(\bm \lambda_{r}^*)
    d\bm z^*_r d\bm \lambda^*_r \times \\
    & \int_{\mathcal S_{\bm \gamma^*_r, \bm \lambda_r^*}} p(\bm{\hat v} | \bm \gamma_r^*, \lambda^*_r) p(\lambda_r^*) p(\bm \gamma^*_r) d\lambda^*_r
    d\bm \gamma^*_r
  \end{split}
  \label{marginal_lkh}
\end{equation}

The quantity $L(\bm y^*|\mathcal M_r)$ (also referred to as marginal likelihood of model $\mathcal M_r$) is usually estimated with Monte Carlo
techniques (\cite{Vyshemirsky2008}).  In this study Annealed Importance Sampling is used. It is important to notice that, because of the adopted prior
probability distributions for the hyper parameters in $\bm \gamma^*_r$, inadequate models with $\bm x_s$ highly correlated with the discrepancy, will
have low $L(\bm y^*|\mathcal M_r)$. Model selection can then be performed according to Table \ref{TAB:BF_scale}.
\begin{table}
  \caption{\label{TAB:BF_scale}Reference values for Bayes Factor interpretation.}
  \centering \fbox{
    \begin{tabular}{*{3}{c}}
      $log_{10}(\mathcal B_{j,i})$ & $\mathcal B_{j,i}$ & Evidence\\
      \hline
      $< 0$      & $< 1$       & negative\\ 
      $[0, 1/2]$ & $[1, 3.2]$  & weak\\
      $[1/2, 1]$ & $[3.2, 10]$ & substantial\\
      $[1, 2]$   & $[10, 100]$ & strong\\
      $> 2$      & $> 100$     & decisive\\
    \end{tabular}
  }
\end{table}

In order to perform a sensible model comparison it is also advised to follow two sensible rules: (i) to compare models built upon simulation samples
of the same size, and (ii) to keep the prior probability density distributions unchanged with respect to rigid translation of the parameter space.

Ideally the model selection process would stop when all the boundary conditions have $\tilde{\mathcal R}^2$ indexes not significantly different from
zero. However, in real cases this is rarely going to be the case, and the analysis can terminate when the last retained model reaches a sufficient
prediction accuracy, as measured by the RMSE. However it is important to highlight that in the presented approach RMSE is not used to performed model
selection, which is undertaken according to Bayes Factors, but only as indication of the likely error embedded in the predictions of the retained
model.

\section{Experiments and results}
\begin{figure}
  \centering
  \begin{minipage}[t]{0.4\textwidth}
    \centering \makebox{\includegraphics[width=\textwidth]{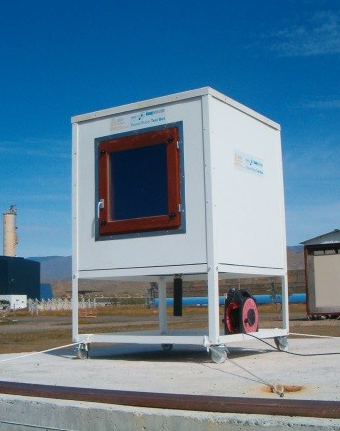}}
    \caption{\label{FIG:testBox_BBRI_site}The box at the test site.}
  \end{minipage}
  \begin{minipage}[t]{0.5\textwidth}
    \centering \makebox{\includegraphics[width=\textwidth]{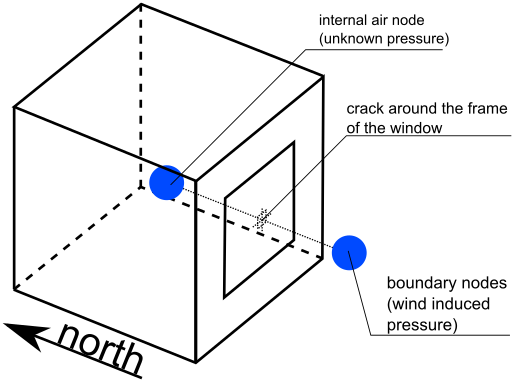}}
    \caption{\label{FIG:testBox_model}Graphical representation of model $\mathcal M_3$.}
  \end{minipage}
\end{figure}

The test facility (box) is shown in Figure \ref{FIG:testBox_BBRI_site}, and it had cubic form with internal dimensions $96 \times 96 \times 96$ cm.
The roof, floor and walls had all identical, but unknown, composition and thickness of 12 cm.  The front wall had a window of dimensions
$60 \times 60$ cm, wherein the glazed part had an area of $52 \times 52$ cm.  The whole structure was provided with a support which allowed the
influence of the ground to be neglected during the modelling.

The box was used in round robin experiments in the context of IEA Annex 58. In the present paper the focus in on two tests led by the Building
Component Energy Test Laboratory (LECE), in Almeria (Spain). The first test (also referred to as ROLBS test) lasted four days (28/06/2013--1/07/2013),
and consisted of monitoring the internal temperature and the heat flux due to conduction through the walls, while applying a Randomly Ordered
Logarithmic Binary Sequence (ROLBS) of heat pulses \citep{Dijk1995} to the box. In the second test (also referred to as free-float test), the internal
temperature was left free to vary, according to the solicitations induced by the weather, and monitored for a period of eight days
(02/07/2013--10/07/2013). The boundary conditions were monitored and provided as well.  These consisted of: external temperature ($Te$), global
vertical solar radiation of the window plane ($Gv$), wind speed ($Ws$), wind direction ($Wd$) and ROLBS heating pulses ($RHP$). For more details the
reader is referred to \cite{JimenezTaboada2016}. The measurement time step was 1 minute, resulting in time series of 5760 and 11520 observations
respectively for the ROLBS and free-float test. Thus, the data were post-processed by bootstrapping the raw time series so as to calculate 15-minutes
averages. This allowed a rough estimation of the measurements standard errors, which were used to set-up the observation precision prior probability
density distributions.

The calibration experiments undertaken consisted of calibrating a series of six models against the box internal temperature measured during the ROLBS
test. These models were created according to the results provided by the discrepancy analysis and their main characteristics are described in the
following. The conduction heat flux observed during the ROLBS test, and the internal temperature observed during the free-float test, were used for
validation purposes.

Before describing the results of the calibration experiments against the observed internal temperature (real experiments), a series of analogous
synthetic experiments, for which the true solution was known and considering only the first three models, is presented in order to prove the
capability of the method for identifying suitable upgrades, leading to the more adequate model among the built variants.

\begin{table}[!]
  \centering
  \caption{\label{TAB:pars}Parameter prior probability density distributions. x: present in the model, /: not present in the model.}
  \fbox{
    \begin{tabular}{*{9}{c}}
      Parameter & Initial value & Distribution & $\mathcal M_1$ & $\mathcal M_2$ & $\mathcal M_3$ & $\mathcal M_4$ & $\mathcal M_5$ & $\mathcal M_6$\\ 
      \hline
      \begin{tabular}{c}$wall_k$\\($W/mK$)\end{tabular} & NA &$\mathcal Unif(0.07, 0.13)$ & x & / & / & / & / & / \\ 
      \begin{tabular}{c} $wall_c$\\($kJ/m^3K$)\end{tabular} & NA & $\mathcal Unif(1680, 3120)$ & x & / & / & / & / & / \\
      $window_t$ (-) & NA & $\mathcal Uniform(0.5, 0.72)$ & / & / & / & x & / & / \\
      \begin{tabular}{c} $crck1_A$\\($mm^2$)\end{tabular} & 790 & $\mathcal Unif(700, 1300)$ & / & /& x & / & / & / \\
      \begin{tabular}{c}$crck2_A$\\($mm^2$)\end{tabular} & NA &  $\mathcal Unif(10, 610)$ & / & / & / & / & x & x \\ 
      \begin{tabular}{c}$wall_{ext,k}$\\($W/mK$)\end{tabular} & 1.05 & $\mathcal Unif(0.7, 1.3)$ & / & x & x & x & x & x \\
      \begin{tabular}{c}$wall1_{ext,c}$\\($kJ/m^3K$)\end{tabular} & 3361 & $\mathcal Unif(2240, 4160)$ & / & x & x & x & x & / \\
      \begin{tabular}{c}$wall2_{ext,c}$\\($kJ/m^3K$)\end{tabular} & NA & $\mathcal Unif(1440, 3360)$ & / & / & / & / & / & x \\ 
      \begin{tabular}{c}$wall_{ins,k}$\\($W/mK$)\end{tabular} & 0.048 & $\mathcal Unif(0.035, 0.065)$ & / & x & x & x & x & x \\ 
      \begin{tabular}{c}$wall_{ins,c}$\\($kJ/m^3K$)\end{tabular} & 179 & $\mathcal Unif(112, 208)$ & / & x & x & x & x & x \\ 
      $r/c$ (-) & 0.79 & $\mathcal Unif(0.53, 0.98)$ & x & x & x & x & x & x \\
    \end{tabular}
  }
\end{table}

\subsection{Models}
Table \ref{TAB:pars} shows the considered parameters for the different models with the respective prior probability density distributions.

Model $\mathcal M_1$ was built in order to be the simplest representation of the test facility.  The box was considered completely sealed (i.e. no air
infiltration from the outside was allowed).  Since the real composition of the walls was unknown, they were approximated with construction components
having only one material layer.  The same construction was also used for the window frame. For the glazed component a set of standard fixed property
values was assumed.  The heating system providing the $RHP$ was considered perfect (i.e. with no heat capacity of its own and able to instantaneously
transmit the heat to the test box walls and internal air). The set of considered model parameters consisted of: wall conductivity ($wall_k$), wall
heat capacity ($wall_c$), and the coefficient determining the fractions of radiative and convective heat provided by the heating system ($r/c$).

Model $\mathcal M_2$ upgraded $\mathcal M_1$ by substituting the the single material layer construction with one composed of three material layers:
two external layers of thickness 3 cm ($wall_{ext}$), made of the same material with high heat capacity and relatively high conductivity, enclosing a
third layer of thickness 6 cm ($wall_{ins}$) with low heat capacity and low conductivity.  Consequently the conductivities ($wall_{ext,k}$ and
$wall_{ins,k}$) and heat capacities ($wall1_{ext,c}$ and $wall_{ins,c}$) of the new construction component replaced the conductivity and density of
the old one ($wall_k$ and $wall_\rho$).

Model $\mathcal M_3$ was built by adding to $\mathcal M_2$ a component modelling a small crack around the window frame, hypothesising that the window
was not perfectly sealed thus allowing the infiltration of air from the outside. The area of the crack was added to the calibration parameter set
($crack1_A$).

Model $\mathcal M_4$ consisted in upgrading $\mathcal M_2$ by better modelling the glazed component and adding the glass optical transmission
($window_t$) to the calibration parameters.

Models $\mathcal M_5$ and $\mathcal M_6$ had the same structure of $\mathcal M_3$, but the model parameter space was rigidly translated so as to
change the ranges of $crack1_A$ and $wall1_{ext,c}$. In $\mathcal M_5$ the parameter $crack1_A$ was substituted with $crack2_A$. In $\mathcal M_6$ the
parameters $crack1_A$ and $wall1_{ext,c}$ were replaced with $crack2_A$ and with $wall2_{ext,c}$ respectively.

\subsection{Synthetic Experiments}
Model $\mathcal M_3$ was considered as representing the true box, and synthetic observations were generated by running it with parameters set to the
initial values displayed in Table \ref{TAB:pars}, and adding white noise to the simulation output according to a noise to variance ratio of 0.01. In
this way inadequacies were introduced in models $\mathcal M_1$ and $\mathcal M_2$, while model $\mathcal M_3$ was free from this kind of errors. These
inadequacies were as follows.

Model $\mathcal M_1$, presented two deficiencies with respect to the true model. Firstly, the lack of heat capacity due to its single layer
construction caused a faster transmission of heat provided by $RHP$ to the internal air. This made its internal temperature excessively sensitive to
the heat pulses. Secondly, due to the absence of wind driven air infiltration from the outside, the internal temperature was less affected by $Te$,
$Ws$, and $Wd$.

Model $\mathcal M_2$, being provided with the correct multi-layer construction, had only the second of the above explained inadequacies.

\begin{figure}
  \centering \fbox{\begin{tabular}{ccc}
                     \includegraphics[width=0.30\textwidth]{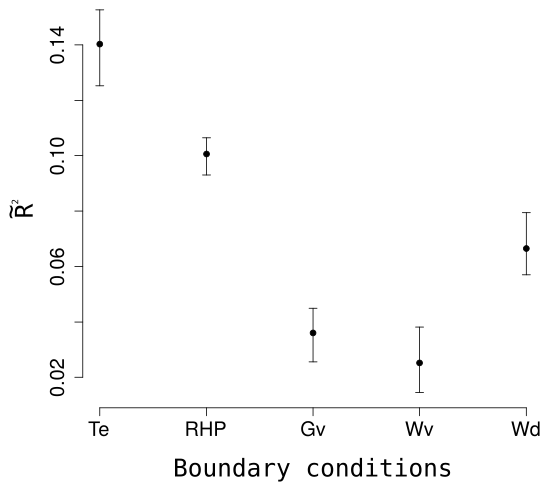} & \includegraphics[width=0.30\textwidth]{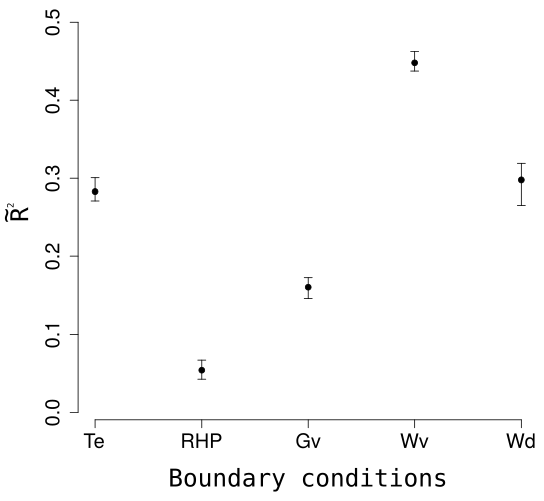} & \includegraphics[width=0.30\textwidth]{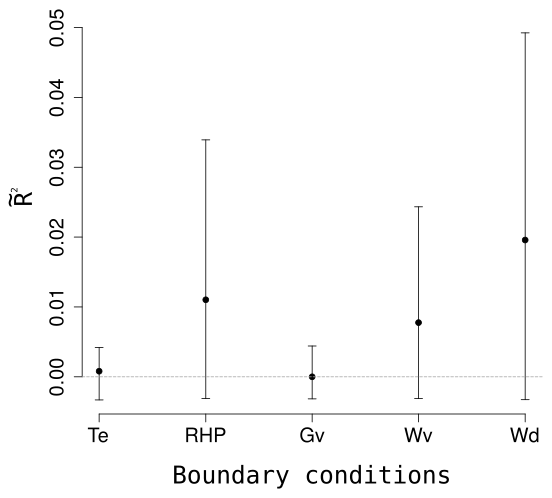}\\
                     (a) model $\mathcal M_1$. & (b) model $\mathcal M_2$. & model $\mathcal M_3$. \\
                   \end{tabular}}
                 \caption{\label{FIG:synth_DA}Synthetic experiment: discrepancy analysis results. The whiskers indicate the 95\% high density
                   intervals.}
               \end{figure}

               The plots in Figure \ref{FIG:synth_DA} show the values of the $\tilde{\mathcal R}^2$ indexes of the considered boundary conditions for
               the three models. The whiskers indicated the 95\% high density intervals. With respect to model $\mathcal M_1$ all the boundary
               conditions were significant. However $Te$ and $RHP$ appear to have an higher importance, clearly meaning a lack of heat
               capacity. Similar results were obtained for, model $\mathcal M_2$, but this time $Ws$ and $Wd$ had the highest $\tilde{R}^2$,
               highlighting the possibility of wind driven infiltration. For model $\mathcal M_3$ all the $\tilde{\mathcal R}^2$ indexes were not
               significantly higher than zero, confirming that this model did not have any inadequacy.

               The results in Table \ref{TAB:synth_calres} give further support to the observation just made. Table \ref{TAB:synth_calres}a lists the
               models' marginal likelihoods, demonstrating that the implemented upgrades, with respect to $\mathcal M_1$, were decisive improvements
               ($\mathcal B_{2,1} = 296.14$ with $HDI = [278.59, 312.06]$, and $\mathcal B_{3,2} = 864.96$ with $HDI = [859.06, 875.37]$). Table
               \ref{TAB:synth_calres}b contains the estimates and high density intervals (HDI) for the parameters of model $\mathcal M_3$. Despite
               minor inaccuracies with respect to weak parameters (e.g. $wall_{ext,k}$ was slightly overestimated), all the variables were accurately
               inferred.

\begin{table}[!]
  \centering
  \caption{\label{TAB:synth_calres}Synthetic experiments: estimates and 95\% HDI for $log_{10}[L(\bm y*|\mathcal M_r)]$, and model $\mathcal M_3$
    parameters.}
  \begin{tabular}{cc}
    (a) $log_{10}[L(\bm y^*|\mathcal M_r)]$. & (b) model $\mathcal M_3$ parameters. \\
    \fbox{\begin{tabular}[t]{*{3}{c}}
            Model & Estimate  & 95\% HDI \\
            \hline
            $\mathcal M_1$ & -348.93 & [-362.19, -340.40] \\
            $\mathcal M_2$ & -52.79 & [-62.58, -49.81] \\
            $\mathcal M_3$ & 812.17 & [809.07, 814.38] \\
          \end{tabular}
    }& \fbox{\begin{tabular}[t]{*{3}{c}}
               Parameter & Estimate & 95\% HDI \\
               \hline
               $crck1_A$ & 790 & [710, 1200] \\
               $wall_{ext,k}$ & 1.21 & [0.72, 1.26] \\
               $wall1_{ext,c}$ & 3443 & [2880, 4077]\\
               $wall_{ins,k}$ & 0.047 & [0.041, 0.056]\\
               $wall_{ins,c}$ & 163 & [117, 202]\\
               $r/c$ & 0.81 & [0.66, 0.91]\\
             \end{tabular}
    }
  \end{tabular}
\end{table}

\subsection{Real experiments}
\begin{figure}[!htb]
  \centering \fbox{\begin{tabular}{ccc}
                     \includegraphics[width=0.30\textwidth]{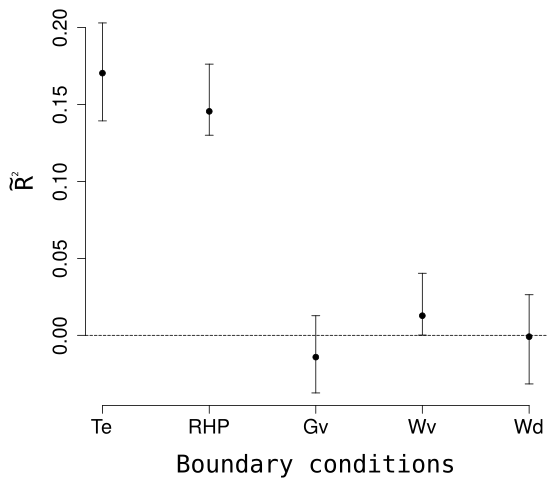} & \includegraphics[width=0.30\textwidth]{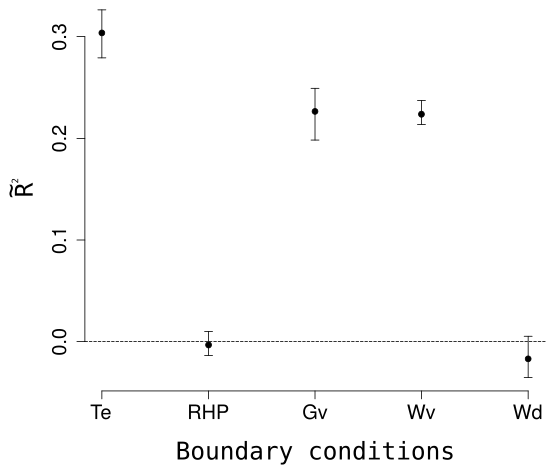} & \includegraphics[width=0.30\textwidth]{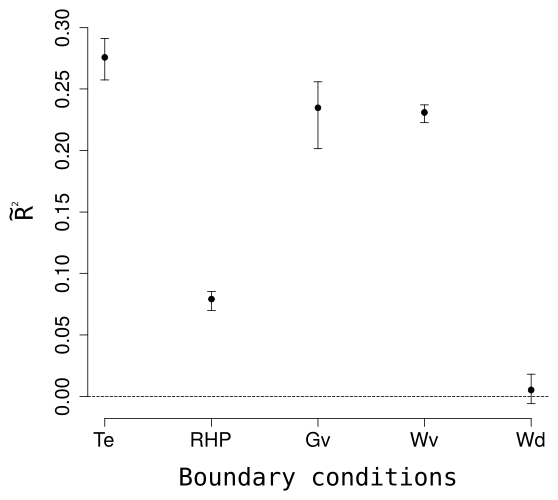}\\
                     (a) model $\mathcal M_1$. & (b) model $\mathcal M_2$. & (c) model $\mathcal M_3$. \\
                     \vspace{1ex}\\
                     \multicolumn{3}{c}{
                     \begin{tabular}{cc}
                       \includegraphics[width=0.30\textwidth]{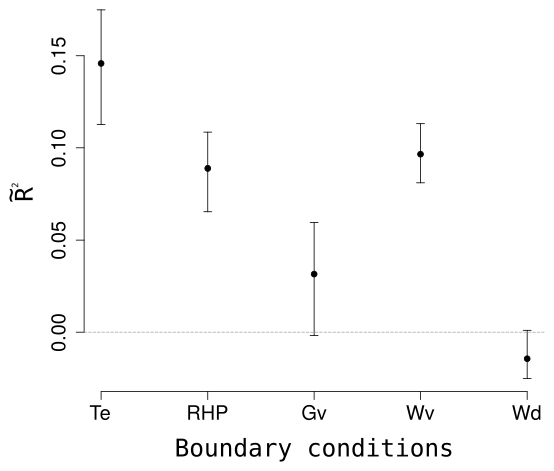} &  \includegraphics[width=0.30\textwidth]{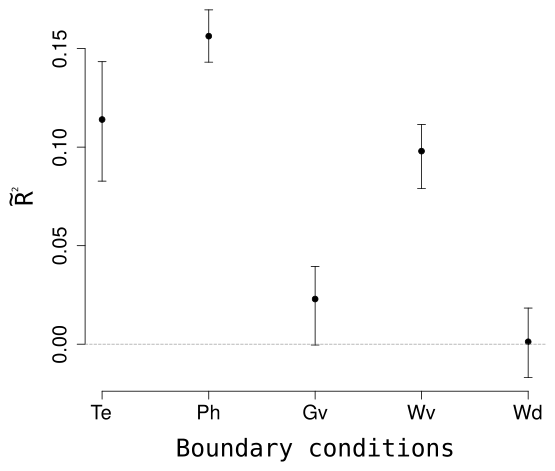} \\
                       (d) model $\mathcal M_5$. & (e) model $\mathcal M_6$. \\
                     \end{tabular}
                     }
                   \end{tabular}}
                 \caption{\label{FIG:real_DA}Real experiment: discrepancy analysis results. The whiskers indicate the 95\% high density intervals.}
               \end{figure}
               Figure \ref{FIG:real_DA} summarises the results of the discrepancy analysis driving the creation of the considered models. Table
               \ref{TAB:real_calres} lists the marginal likelihoods and the Bayes factors, while Table \ref{TAB:real_rmse} contains the RMSEs.

               The investigation started by calibrating model $\mathcal M_1$. The discrepancy analysis provided clear outcomes (Figure
               \ref{FIG:real_DA}a) showing that the ROLBS heating pulses were a dominant factor in determining prediction errors, together with the
               external temperature. This led to the hypothesis that the model was lacking heat capacity. The single-layer construction adopted in
               $\mathcal M_1$ was substituted with a multi-layer one, providing for this deficiency, thus creating $\mathcal M_2$. In particular, the
               two external layers were able to quickly absorb the induced heat gains, while the central layer was assuring that the walls had the
               required thermal resistance.

               The Bayes Factor $log_{10}(\mathcal B_{2,1})$ was estimated as 619.94 (with HDI = [521.67, 715.92]) showing that the proposed upgrade
               was decisive.  Furthermore the implementation of the multi-layer construction reduced to a negligible level the significance of the
               $RHP$ (Figure \ref{FIG:real_DA}b). Nonetheless, other boundary conditions, namely $Te$, $Gv$ and $Ws$, had $\tilde{\mathcal R}^2$
               indexes significantly different from zero. Thus, two variants of model $\mathcal M_2$ were created. $\mathcal M_3$ tried to address
               inadequacies caused by $Te$ and $Ws$, modelling a little crack around the window frame, thus accounting for wind driven
               infiltration. $\mathcal M_4$ focused on deficiencies due to a poor processing of $Gv$, by better modelling the glazed component and
               considering its optical transmission among the calibration parameters.

               Both $\mathcal M_3$ and $\mathcal M_4$ did not show enough evidence for being considered significant improvements upon $\mathcal
               M_2$. However, $log_{10}(\mathcal B_{3,2})$ on average indicated that the consideration of wind driven infiltration was an important
               upgrade. Therefore it was decided to refine the modelling of the crack around the window frame, thus creating $\mathcal M_5$ which
               considered a lower range of values for the area of this feature.

               The comparison between $\mathcal M_5$ and $\mathcal M_2$ yielded a Bayes Factor $log(\mathcal B_{5,2})$ equal to 88.83 (with HDI =
               [54.96, 143.31]), therefore the former was retained as new benchmark model. According to the relative discrepancy analysis results
               (Figure \ref{FIG:real_DA}d) $Te$, $RHP$, $Ws$, were significant. An easy modification was to further adjust the heat capacity of the
               model so as to address issues related to $RHP$. In particular, lower values were considered for the heat capacity of the external
               material layers of the wall construction, thus creating $\mathcal M_6$.

               This last variant had $log_{10}[L(\bm y^*|\mathcal M_6)]$ equal to 186.77 (with HDI = [178.33, 190.06]), resulting in an additional
               decisive model improvement ($log_{10}(\mathcal B_{6,5}) = 20.28$, with HDI = [4.36, 44.60]). Nonetheless, the discrepancy analysis
               returned $\tilde{\mathcal R}^2$ qualitatively unvaried with respect to $\mathcal M_5$. Residual deficiencies were deemed likely to be
               caused by: (i) the real construction still differing from that implemented; (ii) the neglected heat capacity and heat transmission
               delay, properties of the real heating system; and (iii) factors such as the pressure coefficients influencing the magnitude of the wind
               driven infiltration. A better modelling of these aspects was judged impractical due to the lack of the necessary
               information. Furthermore, $\mathcal M_6$ was deemed to be sufficiently accurate (RMSE = 0.67 C) in predicting the internal temperature
               of the box during the ROLBS experiment, and overall an adequate representation of the box. Figure \ref{FIG:real_rolbsFit} displays the
               obtained fit.

\begin{table}[!htb]
  \centering
  \caption{\label{TAB:real_calres}Real experiments: estimates and 95\% HDI for $log_{10}[L(\bm y*|\mathcal M_r)]$ and Bayes factors.}
  \vspace{1ex}
  \begin{tabular}{cc}
    (a) $log_{10}[L(\bm y^*|\mathcal M_r)]$ &  (b) Bayes factors.\\
    \fbox{\begin{tabular}[t]{*{4}{c}}
            Model & Estimate & 95\% HDI \\
            \hline
            $\mathcal M_1$ & -539.30 & [-637.0.2, -503.83] \\
            $\mathcal M_2$ & 77.64 & [26.78, 89.26] \\
            $\mathcal M_3$ & 121.89 & [76.36, 141.00] \\
            $\mathcal M_4$ & 34.89 & [11.27, 43.58] \\
            $\mathcal M_5$ & 166.50 & [145.15, 174.92] \\
            $\mathcal M_6$ & 186.77 & [178.33, 190.06] \\
          \end{tabular}} &
                           \fbox{
                           \begin{tabular}[t]{*{3}{c}}
                             $\mathcal B_{i,j}$ & Estimate & 95\% HDI \\ 
                             \hline
                             $\mathcal B_{2,1}$ & 616.94 & [521.67, 715.92]\\
                             $\mathcal B_{3,2}$ & 44.25 & [-14.48, 91.94]\\ 
                             $\mathcal B_{4,2}$ & -42.77 & [-83.93, 4.95]\\
                             $\mathcal B_{5,2}$ & 88.85 & [54.96, 143.31]\\
                             $\mathcal B_{6,5}$ & 20.28 & [4.36, 44.60] \\
                           \end{tabular}
    }\\
  \end{tabular}
\end{table}

\begin{minipage}[!htb]{0.45\textwidth}
  \centering \captionof{table}{\label{TAB:real_rmse}Real experiments: models' RMSEs against measured internal temperatures.}
  \fbox{\begin{tabular}[t]{*{3}{c}}
          Model & ROLBS & Free float\\
          \hline
          $\mathcal M_1$ & 1.81 & 1.46\\
          $\mathcal M_2$ & 1.41 & 0.84\\
          $\mathcal M_3$ & 1.35 & 0.95\\
          $\mathcal M_4$ & 1.32 & 0.93\\
          $\mathcal M_5$ & 1.23 & 0.93\\
          $\mathcal M_6$ & 0.67 & 0.70\\
        \end{tabular}}
    \end{minipage}
    \hfill
    \begin{minipage}[!htb]{0.45\textwidth}
      \fbox{\centering \includegraphics[width=\textwidth]{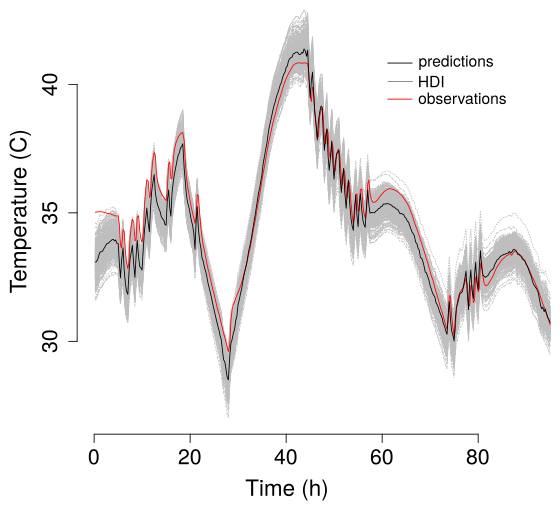}}
      \captionof{figure}{\label{FIG:real_rolbsFit}Real experiment: model $\mathcal M_6$ fit. Measurements: red, model predictions: black, prediction
        95\% HDI: grey.}
    \end{minipage}

    \section{Validation}
    The results obtained from the synthetic experiments demonstrated that the method was able to lead to the progressive identification of the true
    model, which was correctly characterised by the absence of boundary conditions significantly correlated with the discrepancy. With respect to the
    real experiments it was observed that the developed models were able to increasingly better predict the box internal temperature (Table
    \ref{TAB:real_rmse}). However, it was not possible to achieve a model free from inadequacies. Since in this case the true model was unknown, the
    obtained results were validated utilising data not employed during the calibration, and information disclosed after the undertaken analysis. The
    former consisted of the conduction heat flux and the internal temperature measured during the ROLBS and free-float experiments respectively. The
    latter were the real composition of the box walls (Table \ref{TAB:real_wall_cnstr}a), and accurate estimates of its Heat Loss Coefficient (HLC)
    \citep{JimenezTaboada2016}.

\begin{figure}[!htb]
      \fbox{
        \begin{tabular}{cc} \includegraphics[width=0.45\textwidth]{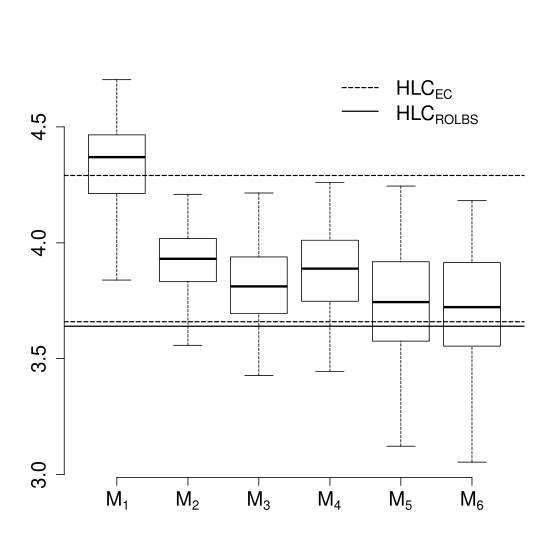} & \includegraphics[width=0.45\textwidth]{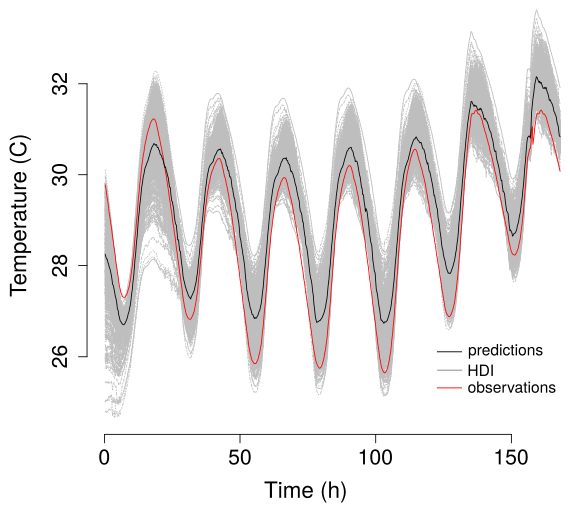}\\
          \begin{tabular}[t]{c}(a) HLC empirical probability
            \\density distributions.\end{tabular} & \begin{tabular}[t]{c}(b) $\mathcal M_6$
                                                      predictions for the free-float
                                                      \\internal temperature.\end{tabular}\\
        \end{tabular}
      }
      \caption{\label{FIG:real_hlcFF}Empirical probability density distribution of HLC obtained through Monte Carlo simulation of the models, and mode
        $\mathcal M_6$ prediction for the box internal temperature during the free float test.}
    \end{figure}

    The HLC, quantifying the heat flowing through the box envelope by conduction for a temperature difference between inside and outside of 1 degree,
    was estimated through a series of tests in an environmental chamber.  Their outcome showed that a sensible range for the HLC was from 3.66 to 4.29
    $W/K$ (also referred to as $HLC_{EC}$ values).  With respect to the ROLBS test, an analogous estimate equal to 3.64 $W/K$ (also referred to as
    $HLC_{ROLBS}$ value) was obtained as the ratio between the total measured heat flux and the total measured temperature difference over the test
    period. The developed computer models were used to estimate these quantities in a Monte Carlo fashion, by sampling the posterior samples inferred
    for their parameters. The results are displayed in Figure \ref{FIG:real_hlcFF}a. The obtained empirical probability density distributions were
    compatible with the range defined by the $HLC_{EC}$ values, especially because the steady-state conditions used in the environmental chamber,
    consisting in elevated temperature differences across the box envelope (up to 50 C), may have caused an overestimation of the HLC with respect to
    the real weather conditions. Indeed, the the HLC estimates returned by the developed models gradually approached the $HLC_{ROLBS}$ value, until
    becoming very similar to it with $\mathcal M_5$ and $\mathcal M_6$ (3.73 and 3.72 $W/K$ respectively).
    \begin{table}
      \caption{\label{TAB:real_wall_cnstr}Real and modelled composition of the box walls.}
      \centering
      \begin{tabular}{c}
        (a) Real composition (from inside to outside).\\
        \fbox{
        \begin{tabular}{*{4}{c}}
          Layer & Thickness ($mm$) & Conductivity ($\frac{W}{mK}$) & Heat capacity ($\frac{kJ}{Km^3}$) \\ 
          \hline
          fibre cement board 1 & 36  & 0.35 & 1838 \\ 
          xps insulation & 60  & 0.034 & 36 \\
          fibre cement board 1 & 16  & 0.35 & 1838 \\
          fibre cement board 2 & 8  & 0.60 & 1960 \\
        \end{tabular}
        }\\\vspace{1ex}\\
        (b) composition adopted for model $\mathcal M_6$.\\
        \fbox{
        \begin{tabular}{*{4}{c}}
          Layer & Thickness (mm) & conductivity ($\frac{W}{mK}$) & Heat capacity ($\frac{kJ}{Km^3}$) \\
          \hline
          $wall_{ext}$ & 30 & 1.24 & 1891.2 \\
          $wall_{ins}$ & 60 & 0.052 & 171 \\
          $wall_{ext}$ & 30 & 1.24 & 1891.2 \\
        \end{tabular}
        }\\
      \end{tabular}
    \end{table}
    
    Similarly, the created models were used to calculate Monte Carlo estimates of the box internal temperature during the free-float test. The Root
    Mean Squared Errors (RMSEs) between the predicted and measured temperatures are listed in Table \ref{TAB:real_rmse}. As the model was being
    upgraded, its capability to extrapolate predictions for the free-float conditions improved. Model $\mathcal M_6$, deemed the best one among the
    created variants, had the lowest RMSE (0.70 C), and its predictions followed well the box internal temperature observed during the free-float test
    (Figure \ref{FIG:real_hlcFF}b).

    Furthermore, the modifications applied according to the results provided by the discrepancy analysis resulted in a modelled wall composition
    qualitatively close to the real one (Table \ref{TAB:real_wall_cnstr}). Probably, an adjustment of the conductivity of layer $wall_{ext}$ towards
    lower values could have additionally improved the model. However, the distributions of thermal capacity and conductivity across the section of the
    modelled walls appeared a good approximation of the real one.

    A final remark is made about the adoption of Bayes Factors as the model selection criterion. The standard practice in the selection of computer
    models is to use a function of the RMSE as measure of goodness and adequacy of the model. If this same principle had been used for comparing the
    models developed, it would not have been easy to choose between $\mathcal M_3$ and $\mathcal M_4$, and probably the latter would have been
    retained for the marginally lower RMSE (Table \ref{TAB:real_rmse}). However, as the analysis has demonstrated, this would have led to the
    inclusion of unnecessary parameters such as the glazed component optical transmission, and to a model having an overestimated HLC (Figure
    \ref{FIG:real_hlcFF}a). Conversely, Bayes factor were always able to provide clear information about model ranking, overall demonstrating more
    robustness with respect to model equifinality.
    
    Given these considerations, the proposed approach was considered reliable and able to provide adequate information leading the modelling activity
    progressively towards a model ($\mathcal M_6$) having properties consistent with the subsequently provided specifications, and capable of
    extrapolating predictions for boundary conditions different from those considered during its calibration.

\section{Conclusions}
The paper has presented an approach for the development of computer models with high-dimensional outputs and boundary conditions, aimed at providing
adequate support for informed decisions driving the modelling activity, in a similar fashion as in statistical modelling. It consisting of: (i) model
calibration in a quasi-Bayesian framework, (ii) the identification of boundary conditions as sources of inadequacy, and (iii) the undertaking of model
selection according to Bayes Factors.

The proposed method was demonstrated on synthetic and real experiments, involving building energy models of a test facility employed for round robin
tests in the context of the IEA EBC Annex 58. However it is applicable to any models presenting high dimensional outputs and boundary conditions.

The results from the synthetic experiments have demonstrated that the method is able to lead to the correct true model.  Ideally, this model is found
when no boundary condition shows relevant correlation with the discrepancy between measurements and predictions. For real cases this is rarely going
to be the case. Nonetheless, the results obtained against the measured data clearly showed that the procedure is capable of providing useful
information about model inadequacies, through the practice called discrepancy analysis. The modeller can use this information to propose consequent
model upgrades, according to its domain knowledge. The effectiveness of these upgrades is then evaluated against the evidence provided by the field
data. In particular, the adoption of Bayes Factors as the model selection criterion was particularly effective in indicating which model variants were
most adequate, leading to a final model of the test facility having properties consistent with specifications provided afterwards the analysis had
been undertaken, and able to effectively extrapolate predictions for boundary conditions other than those used in its calibration.

The explained method has been implemented in the open source software Calibro \citep{Monari2017}, which is being applied and tested, in the context of
a major European project \citep{Costola2017}, on more complex and realistic models.

\bibliographystyle{plain}\bibliography{/home/filippo/Dropbox/MyLibrary/BibTex/PhD}

\end{document}